\documentclass[a4paper]{jpconf}
\usepackage{graphicx}
\begin{document}
\title{Thermal nuclear pairing within the self-consistent
quasiparticle RPA}

\author{N. Dinh Dang$^{1,2}$ and N. Quang Hung$^{1,3}$}
\address{$^1$ Theoretical Nuclear Physics Laboratory, RIKEN
Nishina Center for Accelerator-Based Science,
2-1 Hirosawa, Wako City, 351-0198 Saitama, Japan\\
$^2$ Institute for Nuclear Science and Technique, Hanoi, Vietnam\\
$^3$ Institute of Physics, Hanoi, Vietnam}
\ead{dang@riken.jp (N.D.D.), nqhung@riken.jp (N.Q.H.) }

\begin{abstract}
    The self-consistent quasiparticle RPA (SCQRPA) is constructed to
    study the effects of fluctuations on pairing properties in
    nuclei at finite temperature and $z$-projection $M$ of angular
    momentum. Particle-number projection (PNP) is taken into account
    within the Lipkin-Nogami method. Several issues such as the smoothing
    of superfluid-normal phase transition, thermally assisted pairing in
    hot rotating nuclei, extraction of the nuclear pairing gap using an
    improved odd-even mass difference are discussed.
    A novel approach of embedding the PNP SCQRPA eigenvalues
    in the canonical and microcanonical ensembles is proposed and applied
    to describe the recent empirical thermodynamic
    quantities for iron, molybdenum, dysprosium, and ytterbium isotopes.
\end{abstract}

\section{Introduction}

Sharp phase transitions such as the superfluid-normal (SN) or
shape ones are prominent features of infinite systems such as
metal superconductors, ultra-cold gases, liquid helium, etc. They
are well described by many-body theories such as the BCS, RPA or
quasiparticle RPA (QRPA). The situation changes in finite small
systems such as atomic nuclei, where strong quantal and thermal
fluctuations strongly or completely smooth out these sharp phase
transitions. It is well known that the conventional BCS, RPA or QRPA
theories fail in a number of cases in the description of the ground
states as well as excited states of these systems. The reason is
that strong fluctuations invalidate the assumptions, based on which
the main equations of these theories have been derived. Amongst
these assumptions are the Cooper pairs, which violate the
particle-number conservation, and the closely related
quasiboson-approximation (QBA) used in the (Q)RPA, which violates
the Pauli principle between the fermion pairs. These assumptions
cause the BCS and QRPA to break down at a certain critical value
$G_{c}$ of the pairing interaction parameter $G$, below which the
BCS theory only has a trivial solution with zero pairing gap $\Delta
=$ 0. The same is true in the weak coupling region, where
the particle-particle RPA is valid but its solution also breaks
down at $G\geq G_{c}$. Meanwhile, the exact solution of the pairing
problem exposes no singularity at any $G$~\cite{Exact}. Similarly,
at finite temperature $T\neq$ 0, the omission of
quasiparticle-number fluctuations (QNF) within the BCS theory leads
to the collapse of the pairing gap at the critical temperature
$T_{c}$, corresponding to the temperature of the SN phase transition
in infinite systems. Meanwhile, the exact eigenvalues of the pairing
problem embedded in the canonical ensemble (CE) shows a smooth
decreasing pairing energy with increasing $T$ due to thermal
fluctuations incorporated in the CE~\cite{Ensemble}. In rotating
nuclei, strong fluctuations also smear out the Mottelson-Valatin
effect, according to which the pairing gap, existing at zero
angular momentum $M=$ 0, would collapse at a certain critical
angular momentum $M_{c}$. This situation means that, in order to be
reliable, the BCS, RPA, and/or QRPA theories need to be corrected to
include these effects of fluctuations when applied to nuclei, in
particular, the light ones. This is done within the framework of the
self-consistent QRPA (SCQRPA) presented in this work.

\section{Formalism}

We consider the pairing Hamiltonian $H=\sum_{k>0}\epsilon_{k}
\hat{N}_{\pm k}-G\sum_{kk'}\hat{P}_{k}^{\dagger}\hat{P}_{k'}~$, where
$\hat{N}_{\pm k}=a_{\pm k}^{\dagger}a_{\pm k}$ is the particle-number
operator, and
$\hat{P}_{k}=a_{k}^{\dagger}a_{-k}^{\dagger},
\hat{P}_j=(\hat{P}_j^{\dagger})^{\dagger}$ are the pairing
operators. The operators $a_{k}^{\dagger}$ and $a_{k}$ are
respectively the single-particle creation and destruction operators.
This Hamiltonian has been diagonalized exactly in  \cite{Exact}. The
exact partition function is constructed by embedding the exact
eigenvalues into the CE as $Z_{\rm
Exact}(\beta)=\sum_{S}d_{S}\exp({-\beta\varepsilon_{S}^{\rm
Exact}})$~, with the degeneracy $d_{S}=2^{S}$, inverse temperature
$\beta=1/T$, and $S = 0, 2, ... N$ being the total seniority of the
system. Knowing the partition function $Z$, one calculates the free
energy $F$, entropy $S$, total energy ${\cal E}$, heat capacity $C$,
and pairing gap $\Delta$ as $F = -T{\rm ln}Z(T)$, $S = -{\partial
F}/{\partial T}$, ${\cal E} = F + TS$, $C ={\partial{\cal
E}}/{\partial T}$, and $\Delta=[-G({\cal
E}-2\sum_{k}\epsilon_{k}f_{k}+G\sum_{k}f_{k}^{2})]^{1/2}$, where
$f_{k}$ is the single-particle occupation number on the $k$th level obtained by
averaging the state-dependent occupation numbers $f_k^{(S)}$ within
the CE~\cite{Ensemble}.

The SCQRPA theory~\cite{SCQRPA,SCQRPAT} includes a set of
BCS-based equations, corrected by the effects of QNF, namely
\begin{equation}
    \Delta_{k}=\Delta + \delta\Delta_{k}~,\hspace{2mm}
        \Delta = G\sum_{k'}\langle{\cal D}_{k'}\rangle
        u_{k'}v_{k'}~,\hspace{2mm}
        \delta\Delta_{k}=
        2G\frac{\delta{\cal N}_{k}^{2}}{\langle{\cal
        D}_{k}\rangle}u_{k}v_{k}~.
\label{gap}
\end{equation}
\begin{equation}
       N=2\sum_{k}\bigg[v_{k}^{2}\langle{\cal
       D}_{k}\rangle +\frac{1}{2}\big(1-\langle{\cal
       D}_{k}\rangle\big)\bigg]~.
       \label{number}
       \end{equation}
where $u_{k}$ and $v_{k}$ are the Bogoliubov's coefficients,
\begin{equation}
    u_{k}^{2}=\frac{1}{2}\bigg(1
    +\frac{\epsilon'_{k}-Gv_{k}^{2}-\lambda}{E_{k}}\bigg)~,
    \hspace{5mm}
    v_{k}^{2}=\frac{1}{2}\bigg(1
    -\frac{\epsilon'_{k}-Gv_{k}^{2}-\lambda}{E_{k}}\bigg)~,
    \hspace{2mm} E_{k}=\sqrt{(\epsilon'_{k}-Gv_{k}^{2}
        -\lambda)^{2}+\Delta_{k}^{2}}~,
    \label{uv}
    \end{equation}
with the renormalized single-particle energies $\epsilon'_{k}$
    \begin{equation}
        \epsilon_{k}'=\epsilon_{k}+\frac{G}{
        \langle{\cal D}_{k}\rangle}
       \sum_{k'}(u_{k'}^{2}-v_{k'}^{2})\bigg
       (\langle{\cal A}_{k}^{\dagger}{\cal A}_{k'\neq k}^{\dagger}\rangle
       +\langle{\cal A}_{k}^{\dagger}{\cal A}_{k'}\rangle\bigg)~,
       \label{rene}
       \end{equation}
$\langle{\cal D}_{k}\rangle=1-2n_{k}$, the quasiparticle-pair operators
${\cal A}_k^{\dagger} = \alpha_k^{\dagger}\alpha_{-k}^{\dagger}$,
${\cal A}_k = ({\cal A}_k^{\dagger})^{\dagger}$, and $\delta{\cal N}_{k}^{2}\equiv
n_{k}(1-n_{k})$ is the QNF on $k$th level. To avoid level-dependent
gaps $\Delta_{k}$, the level-weighted gap
$\bar{\Delta}_{k}=\sum_{k}\Delta_{k}/\Omega$ ($\Omega$ is the number
of levels) is considered in the numerical results. Because of
coupling to collective vibrations beyond the quasiparticle mean
field, the quasiparticle occupation number $n_{k}$ is not given by a
Fermi-Dirac distribution of free fermions, but is found from the
integral equation~\cite{SCQRPAT}
\begin{equation}
    n_{k}=\frac{1}{\pi}\int_{-\infty}^{\infty}
    \frac{\gamma_{k}(\omega)(e^{\beta\omega}+1)^{-1}}
    {[\omega-E_{k}-M_{k}(\omega)]^{2}+\gamma_{k}^{2}(\omega)}d\omega~,
    \label{nj}
    \end{equation}
    where the mass operator $M_{k}(\omega)$ and the quasiparticle
    damping $\gamma_{k}(\omega)$ are functions of $n_{k}$, SCQRPA
    eigenvalues $\omega_{\mu}$, SCQRPA ${\cal X}_{k}^{\mu}$ and ${\cal
    Y}_{k}^{\mu}$ amplitudes, SCQRPA phonon occupations numbers
    $\nu_{\mu}$, as well as $u_{k}$ and $v_{k}$. The SCQRPA
    submatrices $A$ and $B$ contain the screening factors
    $\langle{\cal A}_{k}^{\dagger}{\cal A}^{\dagger}_{k'}\rangle$ and
    $\langle{\cal A}_{k}^{\dagger}{\cal A}_{k'}\rangle$ so that the
    set of SCQRPA equations should be solved self-consistently with Eqs.
    (\ref{gap}), (\ref{number}) and (\ref{nj}) to simultaneously determine
    $\bar{\Delta}$, chemical potential $\lambda$, $n_{k}$,
    $\omega_{\mu}$, ${\cal X}_{k}^{\mu}$ and ${\cal
    Y}_{k}^{\mu}$. To eliminate particle-number fluctuations inherent
    in the BCS theory, the Lipkin-Nogami (LN) particle-number projection
    (PNP)~\cite{LN}
    is applied on top of Eqs. (\ref{gap}) and (\ref{number}).
    The ensuing theory, called the LNSCQRPA theory, has also been extended to include the finite
    $z$-projection $M$ of angular momentum (noncollective
    rotation)~\cite{SCQRPAM}. The set of obtained
    equations is formally the same except that now, depending on the
    single-particle spin projections $\mp\gamma m_{k}$ with $\gamma$
    being the angular velocity, one has two types
    of quasiparticle occupation number, $n_{k}^{\pm}$, so that
    $\langle{\cal D}_{k}\rangle=1-n_{k}-n_{-k}$. At $T=$ 0 and
    $M=$ 0 the SCQRPA theory reduces to its zero temperature and
    non-rotating limit, where $\langle{\cal D}_{k}\rangle =
    1/[1+2\sum_{\mu}({\cal Y}_{k}^{\mu})^{2}]$~\cite{SCQRPA}.

\section{Numerical results and discussions}
\begin{figure}
\begin{center}
     \includegraphics[width=9cm]{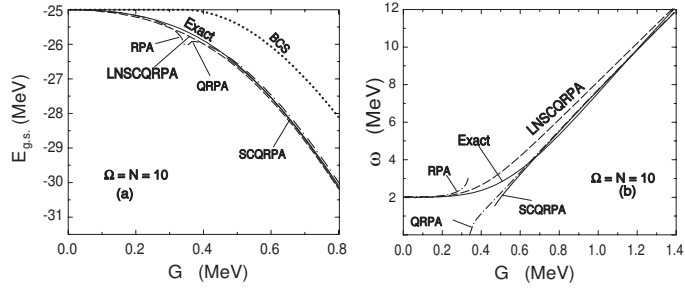}
     \caption{Energies of the ground state (a) and
     first excited states for the $N=\Omega=$ 10 as functions of
     $G$ at $T=M=$ 0. $\omega_{ppRPA}={\cal E}_{1}(N+2) - {\cal E}_{0}(N+2)$
     with the ppRPA eigenvalues ${\cal E}_{i}$.

     \label{EgsE1}}
\end{center}
\end{figure}
Shown in Fig. \ref{EgsE1} are the energies of the ground state (a)
and first excited state (b) obtained at $T=M=$ 0 within several
approximations as well as by exactly diagonalizing the pairing
Hamiltonian for the schematic model, which consists of $\Omega$
doubly-folded equidistant levels with the single-particle energies
chosen as $\epsilon_{k}= k - (\Omega + 1)/2$ MeV. The displayed
results are for the half-filled case with $N=\Omega=$ 10, and
plotted as functions of the pairing interaction parameter $G$. It is
seen that the LNSCQRPA describes rather well the exact energies of
both the ground and first excited states without any discontinuity
in the region around $G_{c}$, where all other approaches such as the
RPA, QRPA, and SCQRPA collapse.

\begin{figure}
\begin{center}
     \includegraphics[width=7cm]{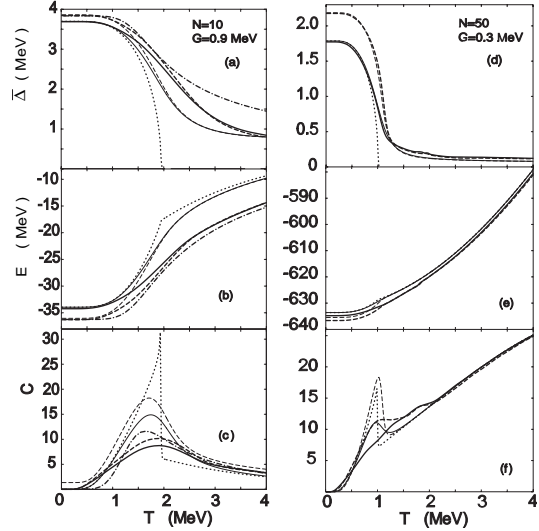}
     \caption{Level-weighted pairing gap $\bar{\Delta}$, total energy $E$, and heat capacity $C$,
     as functions of $T$ for $N=\Omega=$ 10 (a - c) and 50 (d - f) obtained within the
     FTBCS (dotted), FTBCS1 (thin solid), FTLN1 (thin dashed), SCQRPA
     (thick solid), LNSCQRPA (thick dashed). The dash-dotted lines for
     $N=$ 10 are the exact CE results.
     \label{N10N50}}
\end{center}
\end{figure}
The level-weighted gap, total energy, and heat capacity obtained for
the systems with $N=\Omega=$ 10 and 50 are shown as functions of
temperature $T$ in Fig. \ref{N10N50}. Beside
the predictions by the SCQRPA, LNSCQRPA, as well as by their
corresponding limits, FTBCS1 and FTLN1, where coupling to QRPA is
omitted (i.e. $n_{k}$ is described by the Fermi-Dirac distribution
for free fermions), and the finite-temperature (FT) BCS results, the
exact CE results are also shown. This figure clearly demonstrates
how QNF smooth out the sharp SN phase transition in finite systems.
The pairing gap never collapses, but decreases monotonously with
increasing $T$, whereas the spike at $T_{c}$ in the heat capacity,
which serves as a signature of sharp SN phase transition within the
FTBCS, becomes strongly depleted to a broad bump.

\begin{figure}
\begin{center}
     \includegraphics[width=12cm]{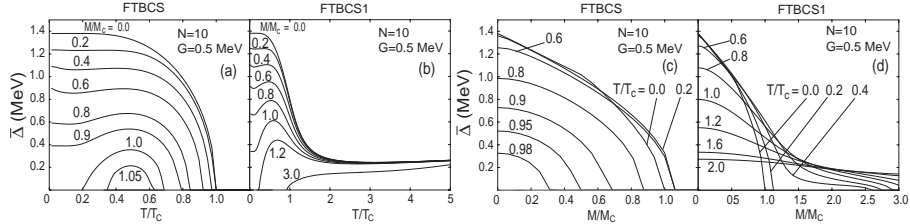}
     \caption{Level-weighted pairing gaps $\bar{\Delta}$ for $N=\Omega=$ 10
     as a functions of temperature $T$ at various values of
     $M/M_{c}$ and angular momentum $M$ at various values of $T/T_{c}$ within the
     FTBCS (a, c) and FTBCS1 (b, d) theories.
     \label{N10M}}
\end{center}
\end{figure}
At finite angular momentum $M\neq$ 0, the FTBCS theory predicts the
Mottelson-Valatin effect, according to which, the zero-temperature
pairing gap decreases with increasing $M$ and collapses at $M =
M_{c}$ because the angular momentum blocks the levels close to the Fermi
surface [Fig. \ref{N10M} (a) and \ref{N10M} (c)]. Thermal effects
relax the blocking, opening some levels around the Fermi surface for
pairing. This leads to the thermally assisted pairing gap (or
pairing reentrance), according to which at a certain $T=T_{1}$ the
pairing gap becomes finite even at $M>M_{c}$~\cite{Moretto,Balian}.
With increasing $T$ thermal effects again break the pairs so that
the gap disappears at $T=T_{2}>T_{1}$ [See Fig. \ref{N10M} (a) for
$M/M_{c}\geq$ 1]. In finite systems, the QNF smooth out both the
Mottelson-Valatin transition and thermal assisted pairing, for
instance, for $N=\Omega=$ 10 with $G=$ 0.5 MeV at $T/T_{c}\geq$ 1,
the gap only decreases monotonously with increasing $M$ but never
vanishes [See Fig. \ref{N10M} (d) for $M/M_{c}\geq$ 1], whereas at
$M/M_{c}\geq$ 3, the pairing gap reappears at $T>T_{1}$ but remains
finite with further increasing $T$  [See Fig. \ref{N10M} (b) for
$M/M_{c}\geq$ 3].

The odd-even mass difference contains the admixture with the
contribution from uncorrelated single-particle configurations, which
increases with $T$. Therefore, the simple extensions of this formula
to obtain the three-point and four-point gaps, in principle, do not
hold at finite temperature. We propose an improved odd-even mass
difference formula at $T\neq$ 0, namely
\begin{equation}
      \widetilde{\Delta}^{(3)}(\beta,N)=
      \frac{G}{2}\bigg[(-1)^{N}+\sqrt{1-4\frac{S'}{G}}\bigg]~,
      \hspace{2mm} S'=\frac{1}{2}\big[\langle{\cal E}(N+1)\rangle_{\alpha}+\langle{\cal
      E}(N-1)\rangle_{\alpha}\big]-\langle{\cal
      E}(N)\rangle_{\alpha}^{(0)}~,
      \label{gapN}
\end{equation}
where $\langle{\cal E}(N)\rangle_{\alpha}$ is the total energy of
the system with $N$ particles within the grand canonical ensemble
(GCE) ($\alpha=$GC) or CE ($\alpha=$C); $\langle{\cal
    E}\rangle_{\alpha}^{(0)}\equiv
    2\sum_{k}\big[\epsilon_{k}-
    Gf_{k}^{(\alpha)}/2\big]f_{k}^{(\alpha)}$ with
$-G\sum_{k}[f_{k}^{(\alpha)}]^{2}$ coming from uncorrelated
single-particle configurations.
        \begin{figure}
        \begin{center}
            \includegraphics[width=12cm]{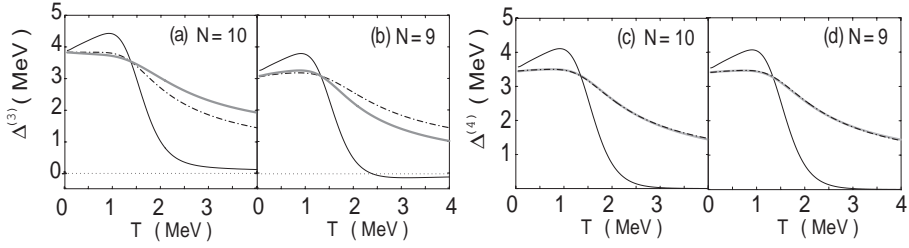}
         \caption{Pairing gaps extracted from the odd-even
         mass differences as functions of $T$ for $N=$ 10 (a,c) and $N=$ 9
         (b,d) ($\Omega=$ 10, $G=$ 0.9 MeV). The thin solid and thick solid lines denote the
         gap $\Delta^{(3)}(N)$
    ($\Delta^{(4)}(N)=[\Delta^{(3)}(N)+\Delta^{(3)}(N-1)]/2)$, and the
    improved gap $\widetilde{\Delta}^{(i)}(\beta,N)$ ($i=$ 3, 4)
    from Eq. (\ref{gapN}), respectively. The dash-dotted lines are the canonical
         gaps $\Delta^{(i)}_{\rm C}$.
         \label{D3}}
         \end{center}
         \end{figure}
    Shown in Fig. \ref{D3} are the pairing gaps $\Delta^{(i)}(\beta, N)$ ($i=$ 3 and
    4), obtained for $N=$ 9 and 10 ($\Omega=$ 10)
    by using the simple extension of the odd-even mass formula to $T\neq$ 0
    as well as the modified gaps
    $\widetilde{\Delta}^{(i)}(\beta,N)$ from Eq. (\ref{gapN}), and
    the canonical gaps $\Delta^{(i)}_{\rm C}$.
    It is seen in Fig. \ref{D3} that the
    naive extension of the odd-even mass formula to $T\neq$
    0 fails
    to match the temperature-dependence of the canonical gap
    $\Delta^{(i)}_{\rm C}$. The gap $\Delta^{(3)}(\beta,N=9)$ even turns negative at $T>$
    2.4 MeV, suggesting that such simple extension of the odd-even mass
    difference to finite $T$ is invalid. The modified
    gap $\widetilde{\Delta}^{(i)}(\beta)$ is found in much better agreement with the canonical
    one, whereas the modified four-point gaps
    $\widetilde{\Delta}^{(4)}(\beta)$ practically coincide with the
    canonical gaps. The
    comparison in Fig. \ref{D3} suggests that formula (\ref{gapN}) is a
    much better candidate for the experimental gap at $T\neq$ 0, rather
    than the simple odd-even mass difference.

    \begin{figure}
    \begin{center}
         \includegraphics[width=12.0cm]{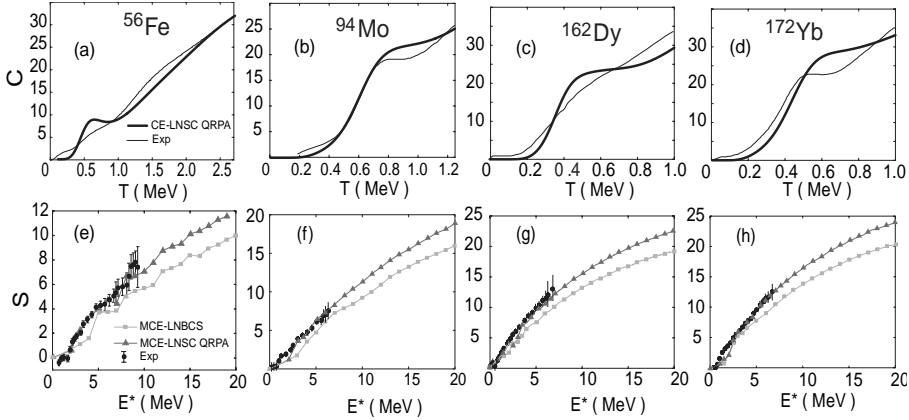}
         \caption{CE heat capacities $C$ as functions of $T$ and MCE
         entropies $S$ as functions of excitation energy $E^{*}$ for
         $^{56}$Fe, $^{94}$Mo, $^{162}$Dy, and $^{172}$Yb.
         Experimental data are taken from \cite{exp}.
         \label{CEMCE}}
    \end{center}
    \end{figure}
    In order to construct a feasible
    description for pairing within the CE, the eigenvalues
    of the LNBCS and LNSCQRPA, obtained for each total seniority $S$ at $T=$
    0, are embedded into the CE
    by using the partition function
    $Z_{\gamma}(\beta)=\sum_{S}d_{S}e^{-\beta\varepsilon_{S}^{\gamma}}$
    ($\gamma=$ LNBCS, LNSCQRPA). The resulting
    approaches are called the CE-LNBCS and CE-LNSCQRPA,
    respectively~\cite{CESCQRPA}. These solutions are also
    embedded into the microcanonical ensemble (MCE) by using the Boltzmann's definition for entropy,
    $S({\cal E})={\rm ln}{W}({\cal E})$, where ${W}({\cal E})$ is the number of
accessible states within the energy interval (${\cal E},{\cal
E}+\delta{\cal E}$). The corresponding approaches are called the
MCE-LNBCS and
    MCE-LNSCQRPA, respectively~\cite{CESCQRPA}.

    The CE heat capacities and MCE entropies for several
    nuclei are shown in Fig.
    \ref{CEMCE} as functions of $T$ and excitation energy
    $E^{*}$, respectively. The single-particle energies are
    calculated within the deformed Woods-Saxon potentials.
    In order to have a consistent comparison with the recent experimental
data in  \cite{exp}, we carried out
calculations by using the CE-LNBCS and CE-LNSCQRPA for
$^{56}$Fe, where pairing is included within the complete $pf+g_{9/2}$
shell above the $^{40}$Ca core. For Mo isotopes, pairing is included
in the 22 degenerated single-particle levels above the $^{48}$Ca
core. For Dy and Yb the same is done on top of the
$^{132}$Sn core. It is clear from this figure that the CE-LNSCQRPA results
    agree quite well with the experimental data~\cite{exp}, which are also deducted from the CE.
    The MCE entropies, obtained within the MCE-LNBCS and MCE-LNSCQRPA
    using ${\delta\cal E} = $1 MeV, are plotted versus the experimental data.
    It is seen that the MCE-LNSCQRPA entropy not only
    offers the best fit to the experimental data
    but also predicts the results up to higher $E^{*} >$ 10 MeV.
\section{Conclusions}

The proposed LNSCQRPA theory can describe without discontinuity the
pairing properties of hot noncollectively rotating nuclei at
any values of pairing interaction parameter $G$, temperature $T$,
and angular momentum $M$. In the limit of zero temperature and zero
angular momentum, it offers the best fits to the exact solutions in
the weak coupling region with large particle numbers for the energy
of the first excited state, whereas the SCQRPA reproduces well the
exact one in the strong coupling region. In the limit of very large
$G$ all the approximations predict nearly the same value as that of
the exact one. Under the effect of QNF, the paring gaps obtained at
different values $M$ of angular momentum decrease monotonously as
$T$ increases, and do not collapse even at hight $T$. The effect of
thermally assisted pairing (pairing reentrance) shows up in such a way that
the pairing gap reappears at a given $T_1 >$ 0 and remains finite at
$T > T_1$, in qualitative agreement with the results of
Ref. \cite{Frau}. We suggest a novel formula to extract the pairing
gap at $T\neq$ 0 from the difference of total energies of
even and odd systems, where the contribution of uncorrelated
single-particle motion is subtracted. Its prediction is found in
much better agreement with the canonical gap than the simple
extension of the odd-even mass formula to $T\neq$ 0. Finally, we
embedded the solutions of the LNBCS and LNSCQRPA into the CE and MCE,
and found that the CE-LNSCQRPA
predictions are in quite good agreements with the exact results as
well as the recent experimental data. The present approach can
also describe simultaneously and self-consistently the
experimentally extracted total energy, heat
capacity, and entropy within both CE and MCE treatments.
It is simple and feasible for the application to larger finite
systems, where the exact matrix diagonalization and/or solving the
Richardson equation are impracticable to find all eigenvalues,
whereas other methods, such as the quantum Monte-Carlo calculations,
are time consuming.


\end{document}